\documentclass[11pt,oneside,a4paper]{article}
\usepackage[T1]{fontenc}
\usepackage{ae} 
\usepackage[a4paper,bottom=2.5cm,top=2.5cm,right=3cm,left=3cm]{geometry} 

\usepackage{url}
\usepackage{floatflt}

\usepackage{ifpdf}
\ifpdf
  \usepackage
  {hyperref}
  \hypersetup{
	colorlinks={true},
	linkcolor={red},
	urlcolor={blue},
	citecolor={red},
    pdftitle={Performance Analysis of Multicast Mobility in a
Hierarchical Mobile IP Proxy Environment},
    pdfauthor={Thomas C. Schmidt, Matthias W\"ahlisch \{schmidt,mw\}@fhtw-berlin.de},
    pdfsubject={Mobile Multicast Communication over IP},
    pdfkeywords={Internet Device Mobility, Mobile IPv6, Multicast, Real-Time Communication, Fast Handover, Hierarchical Mobile IPv6}
  }
    \usepackage[pdftex]{graphicx}
 \else
  \usepackage[dvips]{graphicx}
\fi

\author{Thomas C. Schmidt\textsuperscript{1,2}, Matthias
W\"ahlisch\textsuperscript{2} \\
\small{\{schmidt,mw\}@fhtw-berlin.de} \\
\small{\textsuperscript{1}HAW Hamburg, FB Elektrotechnik und Informatik,
Berliner Tor 7, 20099 Hamburg} \\
\small{\textsuperscript{2}FHTW Berlin, Hochschulrechenzentrum, Treskowallee
8, 10318 Berlin}}
\title{Performance Analysis of Multicast Mobility in a
Hierarchical Mobile IP Proxy Environment\footnote{This work was supported in part by the German Bundesministerium f{\"u}r Bildung und Forschung.}
}
\date{}

\begin{document}
\maketitle
\begin{abstract}
Mobility support in IPv6 networks is ready for release as an RFC, stimulating major discussions on improvements to meet real-time communication requirements. Sprawling hot spots of IP-only wireless networks at the same time await voice and videoconferencing as standard mobile Internet services, thereby adding the request for multicast support to real-time mobility.

This paper briefly introduces current approaches for seamless multicast extensions to Mobile IPv6. Key issues of multicast mobility are discussed. Both analytically and in simulations comparisons are drawn between handover performance characteristics, dedicating special focus on the M-HMIPv6 approach.
\end{abstract}


\section{Introduction}
Mobility Support in IPv6 Networks \cite{rfc-draft-mipv6} \nocite{s-mi6-04} is expected to become a proposed standard within these days. Outperforming IPv4, the emerging next generation Internet infrastructure will then be ready for implementation of an {\em elegant}, transparent solution for offering mobile services to its users. 

At first users may be expected to cautiously take advantage of the new mobility capabilities, i.e. by using Home Addresses while away from home or roaming their desktop 'workspaces' between local subnets. Major scenarios in future IPv6 networks, though, move towards the convergence of IP and 3GPP devices, strengthening the vision of ubiquitous computing and real-time communication. The challenge of supporting voice and videoconferencing (VoIP/VCoIP) over Mobile IP remains, as current roaming procedures are too slow to evolve seamlessly, and multicast mobility waits for a convincing design beyond MIPv6 \cite{sw-rrtam-04}.

Synchronous real-time applications s. a. VoIP and VCoIP
place new demands on the quality of IP services: Packet loss, delay and
delay variation (jitter) in a constant bit rate scenario need careful
simultaneous control. A spoken syllable is about the payload of 100 ms
continuous voice traffic. Each individual occurrence of packet loss above
$1 \%$, latencies over 100 ms or jitter exceeding 50 ms will clearly
alienate or even distract the user. Serverless IPv6 voice or videoconferencing applications s. a. the DaViKo software \cite{cpswm-fwvca-03},\cite{daviko} need to rely on mobility management for nomadic users and applications \cite{swcp-gsvi-03} as well as multicast support on the Internet layer.

Challenges are tightened by multicast-based group communication. In
conferencing scenarios each member commonly operates as receiver and as
sender. A mobile environment consequently needs to cope with the tardy
source specific construction of multicast routing trees.

Our ongoing work is concerned with the design and the analysis of multicast mobility solutions. In the present paper we focus on the essential performance aspects of handovers as applied in the currently available Internet Drafts  Fast Multicast Protocol for MIPv6 \cite{rfc-draft-fmcast} and M-HMIPv6 \cite{rfc-draft-mhmipv6-01}. Following a simple reference model, these aspects are measured quantitatively both, in theory and in simulations.

This paper is organised as follows. In section 2 we discuss related works on multicast mobility and briefly introduce the currently available Internet drafts  M-HMIPv6 and M-FMIPv6. Section 3 formalises evaluation criteria and presents our results,  derived from analytical models as well as stochastic simulations. Conclusions and an outlook follow in section 4.

\section{Multicast Mobility Management}
\subsection{Generals}

Multicast group communication raises quite distinctive aspects within a mobility aware Internet infrastructure: On the one hand Multicast routing itself supports dynamic route configuration, as members may join and leave ongoing group communication over time. On the other hand multicast group membership management and routing procedures are intricate and too slow to function smoothly for mobile users. In addition multicast imposes a special focus on source addresses. Applications commonly identify contributing streams through source addresses, which must not change during sessions, and routing paths in most protocols are chosen from destination to source. 

In general the roles of multicast senders and receivers are quite distinct.
While a client initiates a local multicast tree branch, the source may form
the root of an entire source tree. Hence multicast mobility at the sender side
poses the more delicate problem. 

Three principal approaches to multicast mobility are commonly considered (see \cite{jtcllt-ctatm-03} and \cite{ljrp-mmin-04} for a detailed discussion): Bi-directional Tunneling (BT), Remote Subscription (RS) and Agent-based approaches. The MIPv6 draft proposes BT through Home Agents as a minimal multicast support for mobile senders and listeners. Since Home Agents remain fixed, mobility is completely hidden from multicast routing at the price of triangular paths. In BT the Mobile Node always uses its Home Address for multicast operations. In the contrary, Mobile Nodes following a RS strategy always utilize their link-local addresses, thereby displaying all movements to multicast routing. Routing in turn adapts to optimal paths, on the price of unpredictably slow handovers. A mobile source sending with its link-local address remains immobile with respect to application persistence, as multicast sessions are source address aware. 

A  discussion on various Agent-based approaches is ongoing. Suggestions converge in tunneling multicast traffic through agents at fixed positions within the network. They diverge in the way agents are positioned, in the roles they attain and their interactions with multicast routing protocols. Two aspects we would like to stress in the spirit of our discussion: 

\begin{itemize}
\item Multicast agents should remain in close distance to the mobile multicast member, as extensive tunnels seriously impact network performance.
\item The multicast network components should comply with unicast mobility infrastructure.
\end{itemize}

Agent support has also been proposed to MIPv6 unicasting, to smooth and accelerate handover procedures. A proxy architecture for Home Agents introduces the Hierarchical MIPv6 \cite{rfc-draft-hmipv6} as a counter measure on topological dependencies, when Home Agents are located in far distance. Latency hiding based on predictive handovers has been added to IPv6 mobility by the Fast MIPv6 \cite{rfc-draft-fmipv6} proposal. Handover negotiations in FMIPv6 are performed at access routers, which act as topology aware forwarding agents. 

The two multicast mobility approaches introduced below are built on top of either one of these unicast agent approaches. Little modifications to HMIPv6 resp. FMIPv6 signaling are requested and both proposals remain neutral with respect to multicast routing protocols in use.

\subsection{M-HMIPv6 --- Multicast Mobility in a HMIPv6 Environment}

"Seamless Multicast Handovers in  a Hierarchical Mobile IPv6 Environment (M-HMIPv6)" \cite{rfc-draft-mhmipv6-01} extends the Hierarchical MIPv6 architecture to support mobile multicast receivers and sources. Mobility Anchor Points (MAPs) as in HMIPv6 act as proxy Home Agents, controlling group membership for multicast listeners and issuing traffic to the network in place of mobile senders. 

\begin{figure}
  \center
  \includegraphics[scale=0.6]{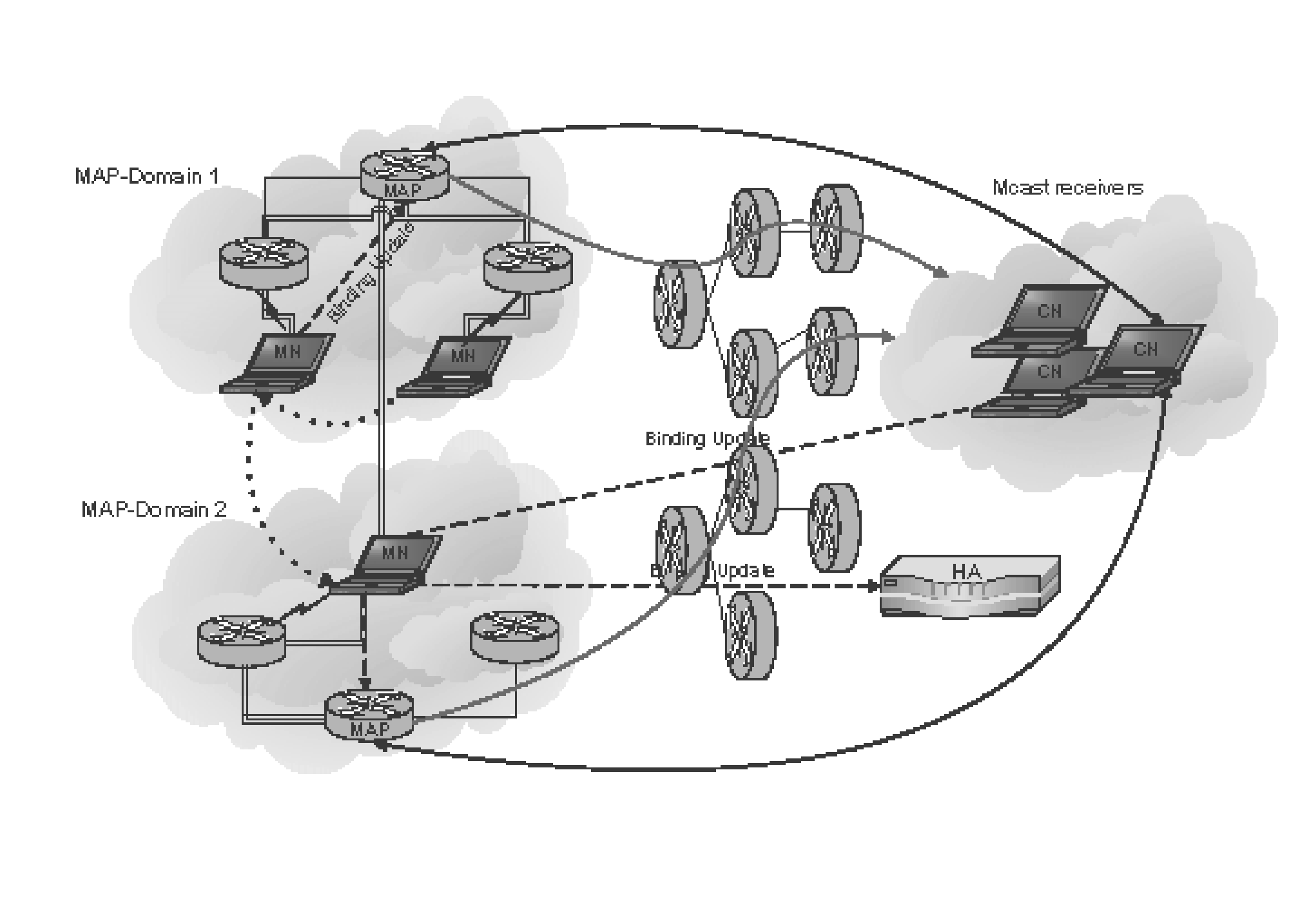}
  \caption{Multicast handover for M-HMIPv6 senders.}
  \label{m-hmipv6}
\end{figure}

All multicast traffic between the Mobile Node and its associated MAP is tunneled through the access network. Handovers within a MAP domain remain invisible in this micro mobility approach. In case of an inter--MAP handover, the previous anchor point will be triggered by a reactive Binding Update and act as a proxy forwarder. A Home Address Destination Option, bare of Binding Cache verification at the Correspondent Node, has been added to streams from a mobile sender. Consequently transparent source addressing is provided to the socket layer.  Bi-casting is used to minimize packet loss, while the MN roams from its previous MAP to a new affiliation (s. fig. \ref{m-hmipv6}). A multicast advertisement flag extends the HMIPv6 signaling.
  
In cases of rapid movement or crossings of multicast unaware domains, the mobile device remains with its previously associated MAP. Given the role of MAPs as Home Agent proxies, the M-HMIPv6 approach may me viewed as a smooth extension of BT through the Home Agent supported in basic MIPv6.

\subsection{M-FMIPv6 --- Multicast Mobility in a FMIPv6 Environment}

\begin{figure}
  \center
  \includegraphics[scale=0.6]{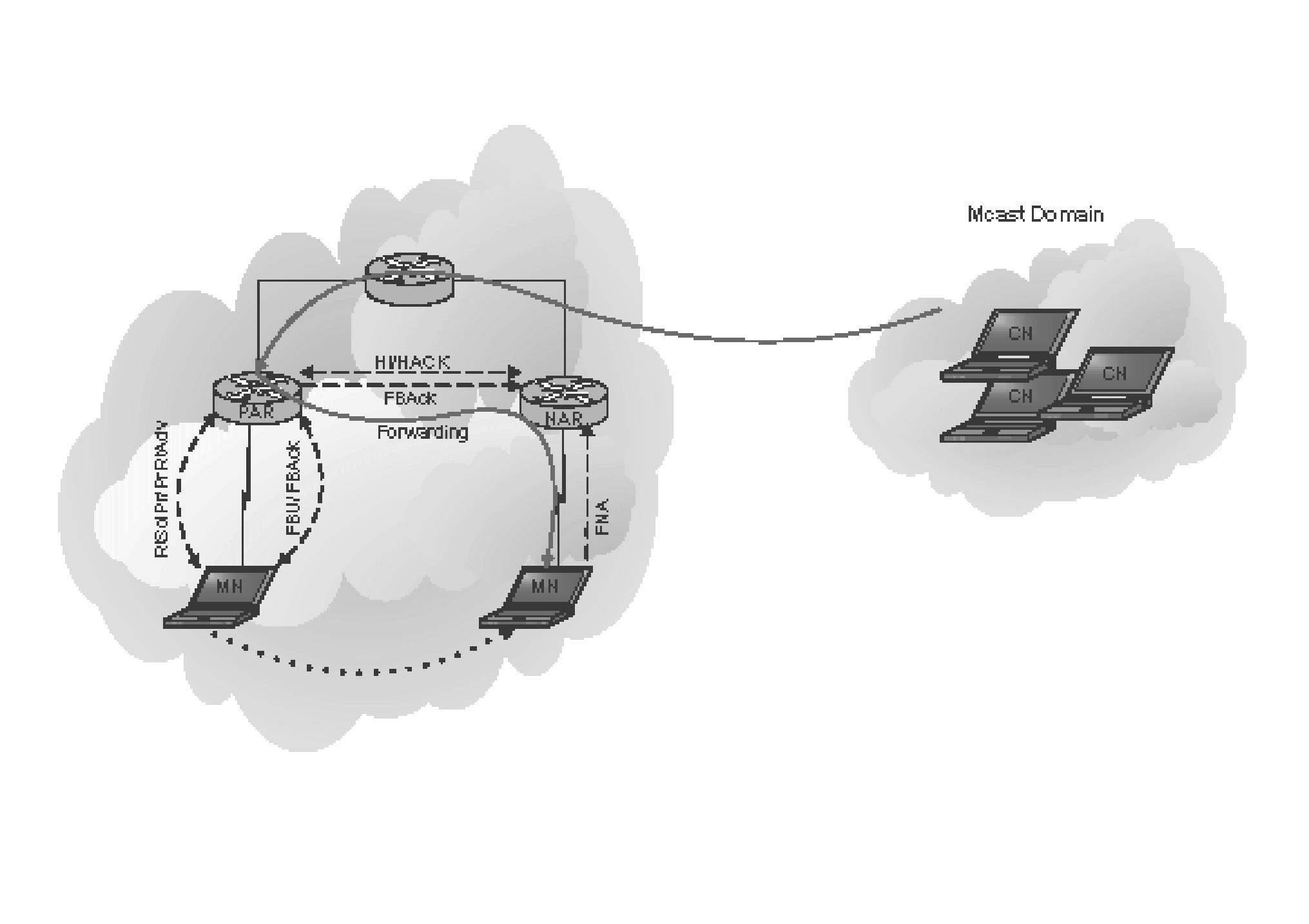}
  \caption{Multicast handover for Fast MIPv6}
  \label{m-fmipv6}
\end{figure}

"Fast Multicast Protocol for Mobile IPv6 in the Fast Handover Environments" \cite{rfc-draft-fmcast} adds support for mobile multicast receivers to Fast MIPv6. On predicting a handover to a next access router (NAR), the Mobile Node submits its multicast group addresses under subscription with its Fast Binding Update (FBU) to the previous access router (PAR). PAR and NAR thereafter exchange those groups within extended HI/HACK messages (s. fig. \ref{m-fmipv6}). In the ideal case NAR will be enabled to subscribe to all requested groups, even before the MN has disconnected from its previous network. To reduce packet loss during handovers, multicast streams are forwarded by PAR as unicast traffic in the FMIPv6 protocol.

Due to inevitable unreliability in handover predictions --- the layer 2 may not (completely) provide prediction information and in general will be unable to foresee the exact moment of handoff --- the fast handover protocols depend on fallback strategies. A reactive handover will be performed, if the Mobile Node was unable to submit its Fast Binding Update, regular MIPv6 handover will take place, if the Mobile Node did not succeed in Proxy Router inquiries. Hence the mobile multicast listener has to newly subscribe to its multicast sessions, either through a HA tunnel or at its local link.  By means of this fallback procedure fast handover protocols must be recognised as discontinuous extensions of the MIPv6 basic operations. 

\section{Analysis of Multicast Handover Schemes}

\subsection{Aspects of Relevance}

Both handover approaches introduced in the previous section attempt to assure the continuous reception of real--time data streams for mobile multicast listeners. M--HMIPv6 uses a reactive handover mechanism, whereas M--FMIPv6 prefers to fulfill a handover prediction, with a fallback to a reactive procedure.

To evaluate these schemes in a comparative analysis, let us first fix the relevant qualitative aspects:

\begin{description}
  \item[Handover performance: packet loss, delay and jitter] occur, while the Mobile Node switches between networks. Packets are lost, while it is disconnected and no packet buffering takes effect. Delay and jitter  are added by the handover procedure, if packets
  are buffered or transmitted through indirect paths.
  \item[Number of performed handovers] determines the frequency of changes between networks, while the Mobile Node moves. 
  \item [Number of processed handovers] denotes the frequency of
  handover procedures processed within the infrastructure. The processed handovers may exceed the actual handover completions, as predictive techniques tend to initiate artificial
   handover procedures.
  \item[Overhead in signalling and traffic distribution] depends on the handover protocols in use and places an extra burden to mobility tasks.
  \item[Robustness] is the expression of dependence of the handover performance
   on changes in network geometry or rapid movement.
\end{description}

The above aspects summarize the amount of disturbance produced by the roaming procedures, the effort of support requested from the networking infrastructure and the overall price to be paid in overhead costs.

\subsection{Handoff Performance}
\label{ho-perf}

\begin{floatingfigure}[r]{55mm}
  \center
  \includegraphics[scale=0.5]{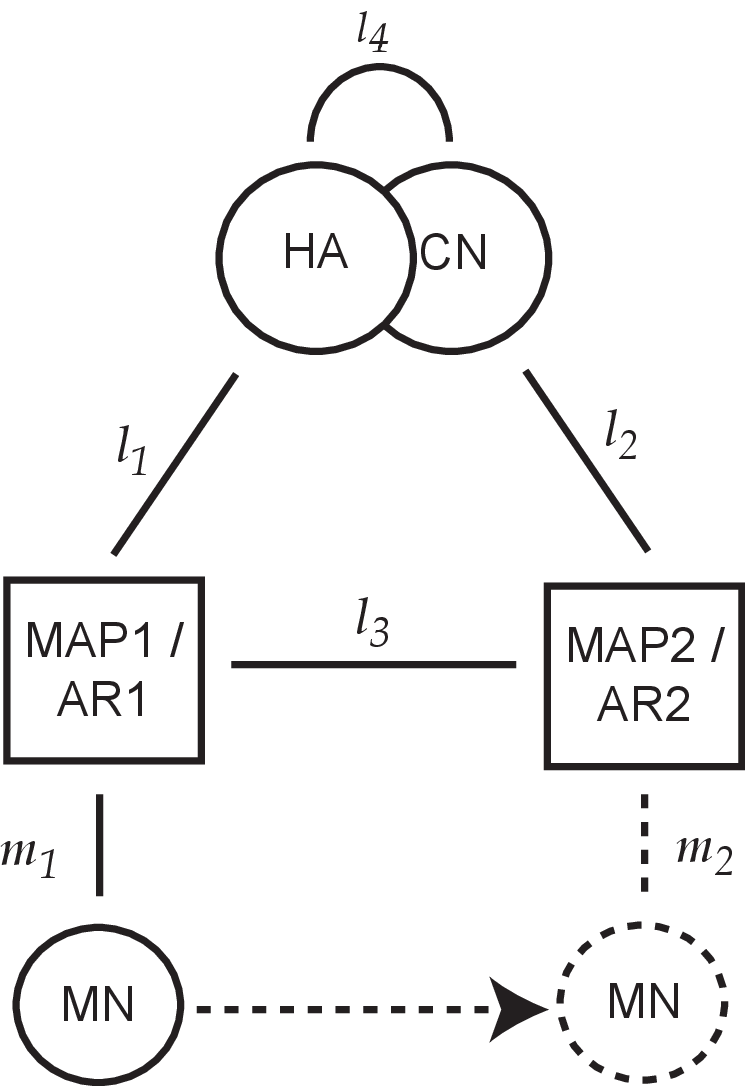}
  \caption{A simple analytical model}
  \label{fanalytmod}
\end{floatingfigure}

In the event of a Mobile Node switching between access networks, it may completely disconnect from the link layer. Thereafter it needs to perform an IP reconfiguration and Binding Updates to its infrastructure. Until completion of all these operations the MN is likely to experience disruptions or disturbances of service, as are the result of packet loss, delay and jitter increases.
In general the handover
process decomposes into the geometry independent local handoff, the Layer 2 link
switching and the IP readdressing, and the geometry dependent Binding Update
activities. Consequently the following decomposition of temporal handover
holds:
\begin{equation}
t_{handoff}=t_{L2}+t_{local-IP}+t_{BU}.
\end{equation}
Both multicast mobility schemes under consideration attempt to optimize the non-local update procedures by means of proxying and anticipating delay hiding techniques.

To proceed into a more detailed analysis of the handover schemes, we
consider a simple analytical model (s. figure
\ref{fanalytmod}) in this section: A MN moves from access router 1 (AR1) to access router 2
(AR2) with intermediate 'link' $l_{3}$.
 From it we will analytically derive basic properties and  present a stochastic simulation of handover performance on its basis. 

Within this model  ARs have been identified with mobility
anchor points 1 and 2 for comparative reasons. This simplification merely
affects the link dimensions $m_{1}/m_{2}$ to the MN, which may not assumed to be of
one-hop-type, but nevertheless small. Distances $l_{1}$ and $l_{2}$ to HA or CN must be viewed as possibly large and represent the strongest topological dependence within the model. The
distance between the access routers, $l_{3}$, should be viewed as a variable,
but characteristic geometric entity. As the MN moves between routers, their
separation represents the gap to be bridged by forwarding, something like
the 'size of the step'.

Let $t_{l}$ denote the transition time of a packet along link $l$, $t_{L2}$
the Layer 2 handoff duration, $t_{local-IP}$ the time for local IP
reconfiguration and $t_{Ant}$ the anticipation period in a predictive
handover, than the following proportional ($\propto$) quantities can be
derived:
\paragraph{Bi-directional Tunneling (MIPv6)}
\[
  \begin{array}{llr}
    \textnormal{Packet loss} & \propto & t_{L2}+t_{local-IP}+t_{m2}+t_{l2} \\
    \textnormal{Additional arrival delay} & \propto & t_{l2}-t_{l1}+t_{m2}-t_{m1}
  \end{array}
\]

\paragraph{Reactive Handover (M--HMIPv6)}
\[
  \begin{array}{llr}
    \textnormal{Packet loss} & \propto & t_{L2}+t_{local-IP}+t_{m2}+t_{l3} \\
    \textnormal{Additional arrival delay} & \propto & t_{l3}+t_{m2}-t_{m1}
  \end{array}
\]

\paragraph{Predictive Handover (M--FMIPv6, successful prediction)}
\[
  \begin{array}{clr}
    \textnormal{Packet loss} & \propto & \Delta^-t+\max(\Delta t+t_{L2}-t_{l3},0)
  \end{array}
\]
\[
  \begin{array}{cl}
    \textnormal{where} & \Delta^\pm t = \max(\pm t_{Ant} \mp 2t_{l3} \mp t_{m1},0) \\
    \textnormal{and} & \Delta t = \Delta^+t - \Delta^-t
  \end{array}
\]
\[
  \begin{array}{llr}
    \textnormal{Additional arrival delay} & \propto & t_{l3}+t_{m2}-t_{m1}
  \end{array}
\]

\paragraph{Predictive Handover (M--FMIPv6, erroneous prediction)}
\[
  \begin{array}{clr}
    \textnormal{Packet loss} & \propto &
    \Delta^+t+\textnormal{reactive handover loss}
  \end{array}
\]

Whereas MIPv6 handover timing is dominated by the update delay with the HA,
predictive and reactive handovers solely depend on the relative geometry of
access routers or MAPs. In the case of a small roundtrip delay (compared to L2
handoff) between ARs, reactive handover admits superior performance,
whereas the predictive approach exhibits faster timing at larger router
distances.

\paragraph{Simulation of Handover Schemes}

For empirical comparison a stochastic simulation of predictive and reactive
handovers was performed according to the setup in our previously described
model. While the HA/CN continuously sends packets in a constant bit rate of
one packet per 10 ms, the MN moves from AR1 to AR2.

\begin{floatingtable}[r]{
  \begin{tabular}{|l|c|c|} \hline
  & $<x>$ & $\xi$ \\ \hline
  Anticipation Time & 50 ms & 30 ms \\ \hline
  L2 Handoff Delay & 50 ms & 10 ms \\ \hline
  $m_{1}, m_{2}$ & 2 ms & 1 ms \\ \hline
  Router Distance & & 2.5 ms \\ \hline
  \end{tabular}}
  \caption{Simulation timers representing mean $<x>$ and variation $\xi$.}
  \label{tsimparam}
\end{floatingtable}

\begin{figure}[h]
  \center
  \includegraphics[scale=0.6]{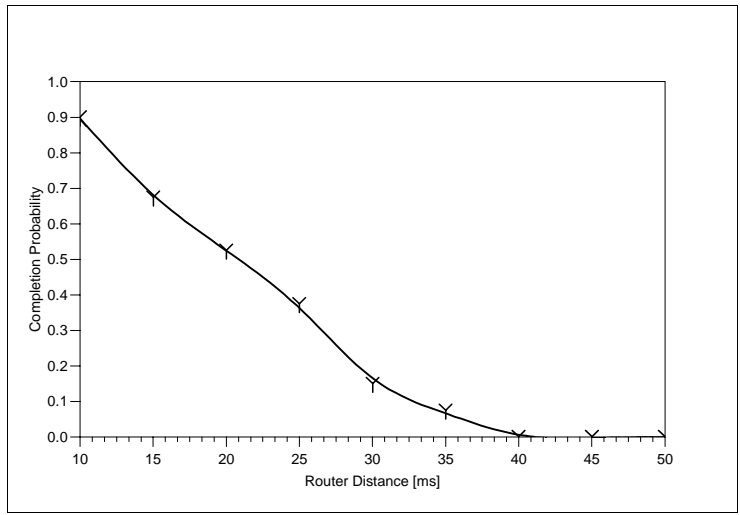}
  \caption{Probability of completing the handover prediction as a function
  of router distance}
  \label{fprob}
\end{figure}

In the case of a predictive handover the MN starts an
anticipation timer and submits a Fast BU to AR1, which subsequently
negotiates with AR2 and eventually returns an FBU acknowledge to the MN. At
this time the MN may have disconnected from its link to AR1. The
probability of successfully completing the anticipation procedure is shown
in figure \ref{fprob} as a function of the router distance ('step size').
As soon as the L2 handoff delay elapsed the MN reconnects with AR2 and is
ready to receive packets forwarded by AR1. Following this procedure we
simulate a 'friendly' predictive handover, i.e. all probabilities of
erroneous predictions as not to arrive in the foreseen subnet, are omitted. For the probability distribution of erroneous predictions see section \ref{ERR-HO-SIM}.

For the reactive handover the MN leaves the link with AR1 instantaneous to
reconnect after the L2 handoff time at AR2. Via AR2 it submits a BU to AR1,
which starts packet forwarding subsequently. 

During the simulation all temporal entities were taken to be random
variables with mean $<x>$ and perturbation $\xi$. The L2 handoff delays were
taken from \cite{sw-rrtam-04}, access link delays at the MN at small scale.
The anticipation period we fixed at the same scale as the L2 handover, but
-- as values are of larger uncertainty -- added large perturbations. In
detail we used the parameter set shown in table \ref{tsimparam}.

\begin{figure}[h]
  \center
  \includegraphics[scale=0.6]{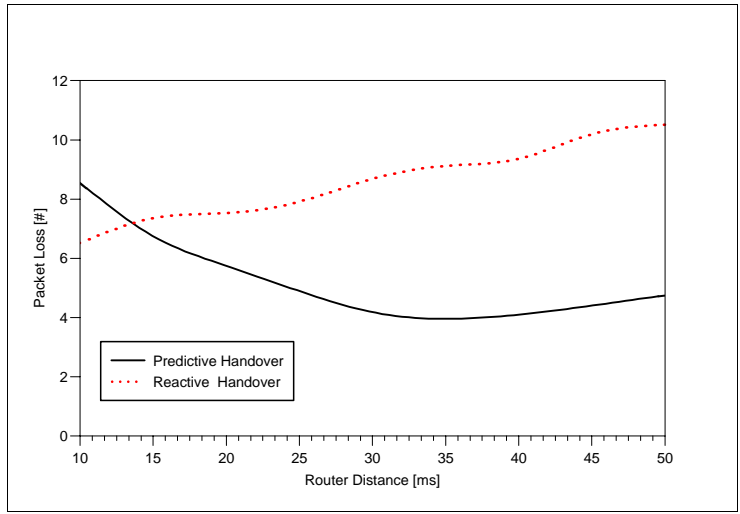}
  \caption{Packet loss in predictive and reactive handover as a function of
   router distance}
  \label{fpredreact}
\end{figure}

Packet loss as functions of router distance are shown in figure
\ref{fpredreact} for predictive and reactive handover simulations
comparatively. Clearly seen can be the prediction artifact of a loss
minimum at a router distance of roughly half the anticipation period. In a
close topology the reactive handover equals or outperforms the predictive
approach, whereas asymptotically its packet loss rate approaches the
prediction results minus layer 2 handoff duration. However, values exceeding a router distance of 25 ms are reached only with strictly limited probability (s. fig. \ref{fprob}).

\subsection{Handover Frequencies}
\label{ERR-HO-SIM}

\paragraph{Expected Number of Handovers}

As a Mobile Node moves handovers potentially impose disturbances and put an extra effort onto the routing infrastructure. Thus the expected frequency of network changes can be viewed as an additional measure of smoothness for a mobility scheme. The handoff frequency clearly depends on the Mobile Node's motion within cell geometry. Two measures on quantizing mobility have been established in the literature: The cell residence time and the call holding time. Both quantities fluctuate according to the overall scenery and the actual mobility event.

Let us make the common assumption that the cell residence time is exponentially distributed with parameter $\eta$ and that the call holding time is exponentially distributed, as well, but with parameter $\alpha$. Then the probability for the occurrence of a handover from MNs residence cell into some neighboring can be calculated analytically to
\[ {\cal P_{HO}} = \frac{1}{1 + \rho}, \  \ where \  \rho = \frac{\alpha}{\eta}\]
is known as the {\em call--to--mobility} factor \cite{fc-agrhp-02}. It can be observed that the handoff probability increases as $\rho$ decreases. Note that all probability distributions are homogeneous in space, e.g. ${\cal P_{HO}}$ is independent of the current cell or the number of previously occurred handovers. Spatial scaling can be applied, accordingly.

\begin{figure}[h]
  \center
  \includegraphics[scale=0.6]{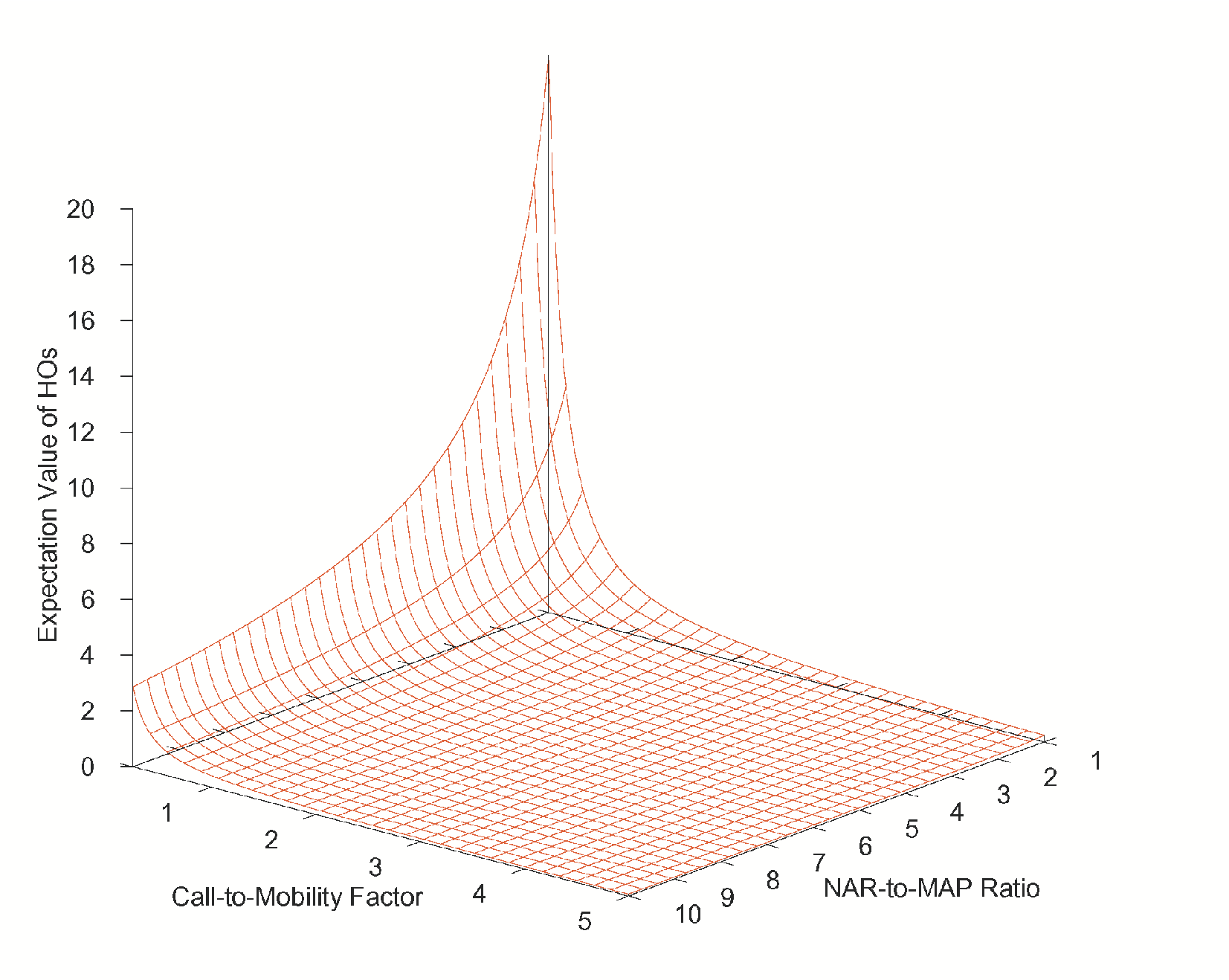}
  \caption{Expectation values for the number of handovers}
  \label{ho-exp}
\end{figure}

When comparing Fast MIPv6 and Hierarchical MIPv6 approaches, another distinctive quantity comes into play: Whereas FMIPv6 operates handovers at Access Routers, HMIPv6 utilizes MAPs, which form a shared infrastructure. In general one MAP is meant to serve $k$ Access Routers, whence the expected number of (inter--MAP) handovers reduces in a HMIPv6 domain. 

Let us assume MAP regions to be of approximately circular geometry. Then the expected cell residence time changes to
\[ \eta^{-1}_{MAP} = \sqrt{k} \cdot \eta^{-1}_{AR} \]
and the handoff probability transforms into

\[ {\cal P_{HO}} = \frac{1}{1 + \sqrt{k} \cdot \rho},\]
where $k$ is the ratio of ARs to MAPs.

Now we are ready to calculate the expected number of handovers as a function of the {\em call--to--mobility} factor $\rho$ and the AR to MAP ratio $k$
\begin{eqnarray}
{\cal E_{HO}} & = & \sum_{i= 1}
^{\infty} i \left( \frac{1}{1 + \sqrt{k} \cdot \rho}\right)^{i} 
  =  \frac{1}{ k \cdot \rho^{2}} + \frac{1}{ \sqrt{k} \cdot \rho}.
\end{eqnarray}
It can be observed that ${\cal E_{HO}}$ increases for decreasing  $\rho $ and $k$ and attains a singularity at $\rho = 0$. Figure \ref{ho-exp} visualizes this expectation value function for common values of $\rho $ and $k$.

\paragraph{Simulation of Prediction Types}

Evaluation of a distribution for handover prediction types cannot be derived analytically. To proceed, we initiated a stochastic simulation of motion within radio cells. This ongoing work combines the following ingredients:

\begin{floatingfigure}[r]{55mm}
  \center
  \includegraphics[scale=0.5]{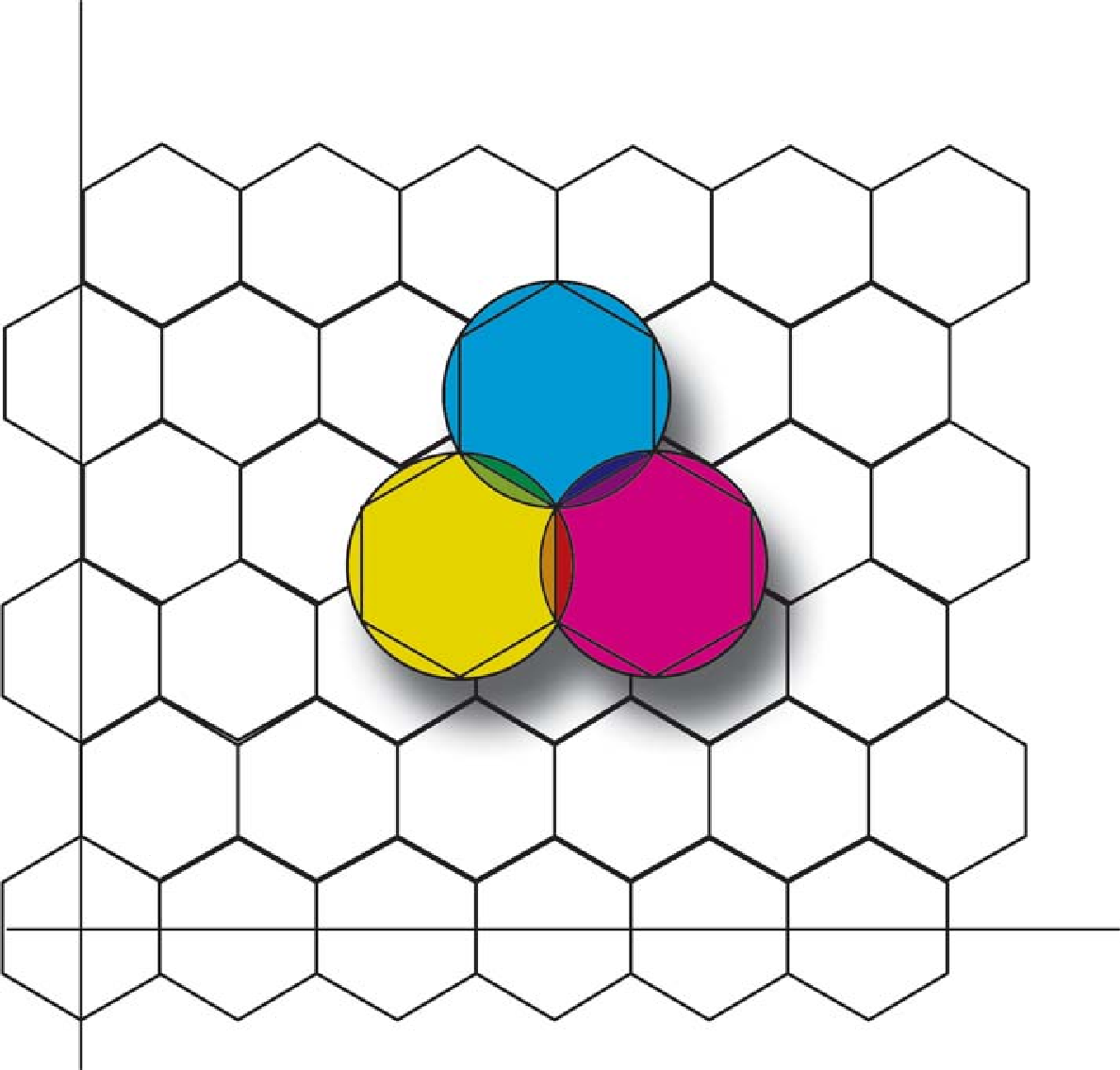}
  \caption{The comb geometry with regions of overlap}
  \label{hcomb}
\end{floatingfigure}

{\em Cell geometry} is chosen to be of honeycomb type, i.e. abutting hexagons completely fill the 2 D plane. The ranges of radio transmission are modeled as minimal circles enclosing the combs. Thus, regions of prediction are the overlapping circle edges as shown in figure \ref{hcomb}. A geometry of coherent Hot Spots is assumed here, where cells -- without loss of generality -- are identified with individually routed subnets.
 
As {\em Walking model} a Random Waypoint Model \cite{b-mmwnc-01} without boundaries was used to be the initial dynamic, i.e. mobile devices move along straight lines for the duration of their (exponentially distributed) call holding time. Boundaries were discarded to avoid border effects. 

Predictions are evaluated along the trajectories, distinguishing their correctness according to the outcome in succeeding steps. In our first simulation experiments a rather strong dependence on mobility contexts appears. Varying the call--to--mobility factor $0 < \rho \leq 5$, we observe an increasing rate of erroneous predictions, ranging from about 1 \% at $\rho = 0.1 $ to exceed 50 \% at $\rho = 5 $. These results account for call termination effects, i.e. the cease of a communication stream, while a handover prediciton is valid. In our current scenario the traversal of prediction regions without subsequent cell change rarely occurs, as may be an expected result in the straight line motion model.

Detailed studies, which quantize effects of geometry and walking models, will be subject to a forthcoming paper.

\subsection{Robustness and Overheads}

A central goal in designing M--HMIPv6 and M--FMIPv6 has been topological robustness in the sense that handover behaviour remains independent of network geometry. Both protocols fulfill this task in a similar way: Any blocking communication with the Home Agent has been excluded. Signalling in both approaches is dominated by the layer 2 handoff and local router exchanges (compare section \ref{ho-perf}).

Another important aspect of robustness must be seen in the ability to cope with rapid movement of the Mobile Node. In the case of a mobile multicast listener leaving its association before a handover completed, an M--HMIPv6 device will remain associated with its previously fully established MAP or Home Agent. On the price of a possible increase of delay, forwarding of multicast data is provided independent of handover frequency. On the contrary M--FMIPv6 forwarding will collapse, as soon as a MN admits a handover frequency incompatible with the signalling completion periods. An M--FMIPv6 device then has to fall back onto MIPv6 by establishing a bi-directional tunnel anew. Meanwhile established services are interrupted.

For final judgment protocol overheads may be taken into account: Both protocols  announce multicast capabilities within regular AR or MAP advertisement. The hierarchical MIPv6 distinguishes cases of handovers between a micro mobile change, which will be accompanied by only one local protocol message, and the inter--MAP handover. The latter is operated by two messages. Handover signalling of fast MIPv6 always requests for seven messages. MIPv6 and multicast routing procedures add on top of both protocols in the event of visible handovers.

\section{Conclusion and Outlook}
Mobility is one of the most challenging and demanded developments in IP
networks today. Mobile multicasting -- as needed in many group conferencing
situations -- places an even stronger challenge on today's IP concepts and
infrastructure. In this paper we presented and analysed the current approaches.

It was found that multicast support based on the fast MIPv6 protocol admits a faster handover at intermediate router distances. Acceleration is reached at the risk of using predictive techniques, which show reduced reliability in the regions of improvement. Starting from simple, fundamental assumptions a quantitative study of expected handover occurrences was derived. The 'nervousness' of handovers performed at access routers could be shown to reduce significantly in the presence of Mobility Anchor Points established within the hierarchical MIPv6 approach. This smoothing effect gains additional importance by observing an instability of fast handovers in the case of Mobile Node's rapid movement.

The exploration of empirical distributions for handover predictions has been started in this ongoing work. Additional modeling and simulations have to be done in order to obtain realistic estimates, while sorting out artifacts of walking models and geometry.


\nocite{s-mi6-04}
\bibliographystyle{IEEEtran}
\bibliography{../Literatur/ipv6}

\end{document}